\documentclass[aps,pre,preprint,groupedaddress]{revtex4-1}

\usepackage{amsmath}   
\usepackage{amsfonts}
\usepackage{graphicx}  
\usepackage{hyperref}  
\usepackage{epstopdf}

\begin{document}

\title{Jeffrey's prior sampling of deep sigmoidal networks}
\author{Lorien X. Hayden}
\author{Alexander A. Alemi}
\author{Paul H. Ginsparg}
\author{James P. Sethna}
\email[]{sethna@lassp.cornell.edu}
\affiliation{Cornell University, Department of Physics, Ithaca, NY 14850}

\date{\today}

\begin{abstract}

Neural networks have been shown to have a remarkable ability to uncover low dimensional structure in data: the space of possible reconstructed images form a reduced model manifold in image space. We explore this idea directly by analyzing the manifold learned by Deep Belief Networks and Stacked Denoising Autoencoders using Monte Carlo sampling. The model manifold forms an only slightly elongated hyperball with actual reconstructed data appearing predominantly on the boundaries of the manifold.  In connection with the results we present, we discuss problems of sampling high-dimensional manifolds as well as recent work [M. Transtrum, G. Hart, and P. Qiu, Submitted (2014)] discussing the relation between high dimensional geometry and model reduction.  

\end{abstract}

\maketitle

\section{Sloppy Models}

Deep neural networks have proven to be state of the art on numerous machine learning benchmark tasks in recent years  \cite{overview}. These networks are so called 'deep' due to their layered structure in which each subsequent layer appears to encode a different, more abstract representation of the data. Previously there has been much interest in the nature of the manifold learned by deep neural networks \cite{BengioMDR13, manifold1, manifold2}.  Through these studies, it has been conjectured that in the higher layers of the network, the representation of the data more uniformly fills the space of neural outputs and that high density regions of the manifold near which raw data concentrates tend to unfold. In other words, linear interpolation between data points in higher layers of the network can be performed without leaving a region of high probability. In contrast, the same procedure in data space would produce a superposition of the two data points which typically results in a point having low probability of naturally occuring in the dataset.  Recently, a method has been developed to approximately perform a Metropolis-Hastings sampling of the data distribution using a trained network\cite{manifold1}. The aim of this method is to generate samples which could be drawn with high probability from the distribution underlying the data. Here we ask a different question, that is, what is the geometry of the manifold the neural network has learned? We present a sampling of the model manifold implementing Metropolis-Hastings weighted by Jeffrey's Prior.  Using this sampling method yields insight into the manifold learned by computational neural networks as well as into Jeffrey's Prior itself. \par

In general, models are created to form a low dimensional representation of the data space. This is true also for neural networks.  In many fields --- including systems biology, statistical physics and mathematical programming --- it has been found that models often form a hierarchical structure in which certain combinations of parameters dominate the behavior while others barely contribute \cite{sloppy1, sloppy2, sloppy3}.  These directions in parameter space are refered to as 'stiff' and 'sloppy' respectively.  Consider a fitting procedure where, given the model and data, the objective is to determine the parameter values. For a least squares cost function, the sloppiness of the model is reflected in the eigenvalues of the Hessian which span many orders of magnitude. Another indicator of sloppiness in a model is that the model manifold, which corresponds to the space of all possible predictions of the model, forms a hyper-ribbon \cite{hyperribbon}. The sloppiest combination of parameters results in the thinnest direction in data space while the stiffest affects the predictions of the model immensely, resulting in a very long direction. \par

Feed forward neural networks have been shown to display the signatures of sloppiness \cite{sloppynn}. In this work, however, we study deep networks whose aim is reconstruction. In particular, we examine Stacked Denoising Autoencoders and Deep Belief Networks trained to reconstruct images from the MNIST dataset, a dataset of handwritten digits in which each pixel takes a value between 0 and 1. In each case, the model manifold forms an only slightly elongated hyperball (Section \ref{sec:results}) which, from the viewpoint of information geometry, indicates the network does well at weighting parameter combinations equally. In addition to this apparent lack of sloppiness,  we also find that the actual data appears predominantly to lie near the boundaries of the manifold; a feature which may be due to the largely saturated pixels of the data set. That most of the images generated by the trained network do not correspond to actual images raises a very interesting question. Does this mean that the neural network is wasting a vast amount of expressive capability? Or does it point to a general feature in modeling that the interesting components of a manifold lie on its edges?  \par

\section{Deep Networks}
The model manifolds we study belong to two types of prototypical deep networks; Deep Belief Networks (DBNs) \cite{HinSal06} and Stacked Denoising Autoencoders (SdAs) \cite{Vincent08}. At their heart, these networks rely on the following mapping between layers of 'neurons' 
\begin{equation} 
    \label{eqn:goup}
	\vec{h}=\sigma(W\vec{d}+\vec{b})
\end{equation}
\begin{equation}
    \label{eqn:godown}
	\vec{y}=\sigma(W'\vec{h}+\vec{b'})
\end{equation}

\noindent where $\vec{d}$ represents the input vector, $\vec{h}$ the hidden activations or output, and $\vec{y}$ the reconstructed input. $W$ and $W'$ are the weight matrices, $\vec{b}$ and $\vec{b'}$ are offset vectors, and $\sigma$ denotes the sigmoid function. By stacking these learned encoding and decoding maps, the network can be trained to compress and reconstruct data.  The map learned by these networks is flexible in that the features in the top hidden layer can be used  as input to a simple classifier such as a softmax layer or support vector machine.\par

DBNs and SdAs differ in the way they are trained. However, both of these types of networks take advantage of a layered structure in which each layer of neurons encodes a different representation of the data. For comparison, we train three of these deep networks on the MNIST dataset \cite{MNIST}. The first network we present is an SdA trained on a single digit, '1'. Since we are interested in networks that perform the reconstruction task rather than classification this is a natural choice to serve as a smaller testbed for our methods. The digit '1' was chosen due to the striking structure observed in PCA plots of the data. We refer to this network as the single-digit network. The other two networks we study are trained on the full dataset and are referred to as the DBN and SdA networks respectively. The particulars for each of the three networks are given in Table \ref{tab:nns}. Each SdA was trained using a cross-entropy loss and stochastic gradient descent via Theano \cite{Theano1,Theano2} and the classes given on deeplearning.net. The DBN was trained using MATLAB code provided by Hinton's group \cite{HinSal06}.\par

\begin{table*}
\begin{center}
\begin{tabular}{| c | c | c | c |}
\hline
& SdA/Single-Digit &  SdA & DBN\\
\hline
data & MNIST '1's & MNIST digits & MNIST digits\\
 training set size& 5678 & 50000 & 50000\\
testing set size & N/A & 10000 & 10000\\
network size &784-100-100-10&784-1000-500-250-30&784-1000-500-250-30\\
class. error & N/A &1.28\%& 1.50\% \\
corruption level & 0.25 & 0.25 & N/A\\
training method & Theano & Theano & Hinton\\
\hline
\end{tabular}
\caption{Training characteristics of the neural networks we study. Classification accuracy on the MNIST dataset \cite{MNIST} was evaluated by training a support vector machine to classify the data given the top layer of features \cite{scikit-learn}. The network and support vector machine were then applied in tandem to the test set in order to calculate the error. Training of the networks were achieved using Theano \cite{Theano1, Theano2} and MATLAB code provided by the Hinton group \cite{HinSal06}. The 4-layer SdA trained with theano was trained with a linear mapping at the top layer. This choice was made to ease comparison between the SdA and the Hinton group's DBN which has this characteristic. For each of the 4-layer networks the layer sizes were chosen to correspond with Hinton's original network. The top hidden layer in each case had a dimension of  30. }
\label{tab:nns}
\end{center}
\end{table*}

Each of these neural networks can be viewed as a fitting process in two distinct ways. For one, training the network is a fit.  During training of the neural network, there are a large number of parameters ($W$, $W'$, $b$ and $b'$ for each layer) and the objective is to fit this multi-parameter model to a large number of images. The manifold associated with this fitting process has been shown to form a ribbon-like structure in high dimensional space in which the manifold becomes thinner and thinner in each successive dimension by a roughly constant factor \cite{sloppynn}. This structure we refer to as a hyper-ribbon. \par

\begin{figure}
\begin{center}
\includegraphics{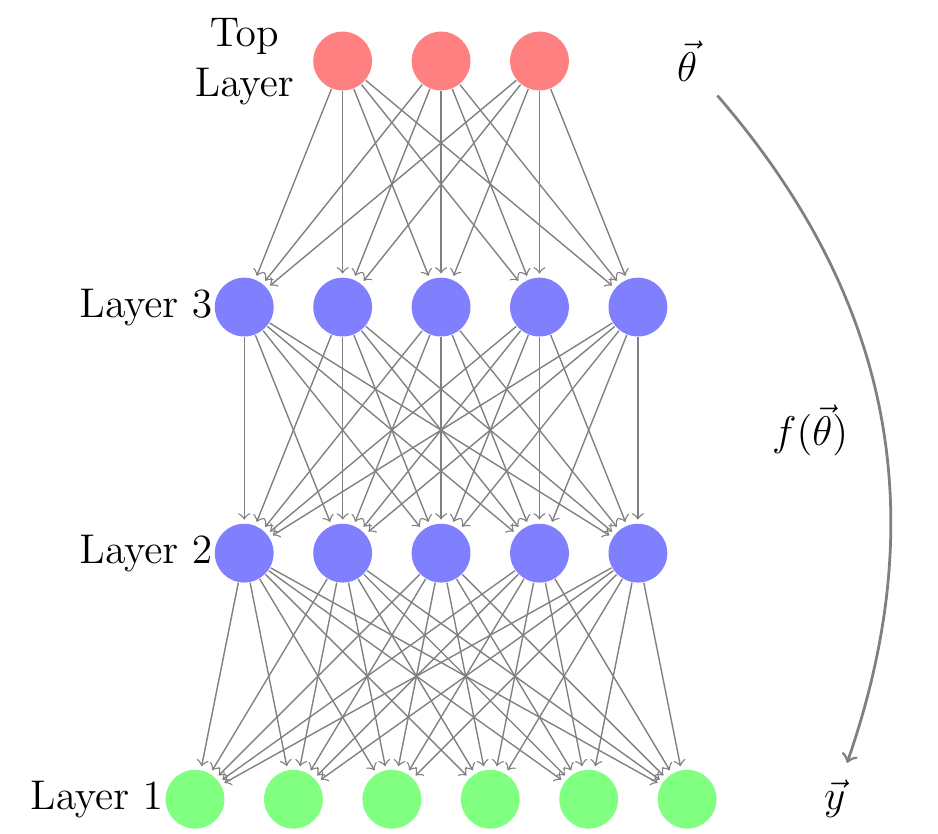}
\caption{(Color online) Diagram of the decoding map learned by a generic 3-layer neural network. Once trained, a DBN or SdA provides a function $f$ such that, for any given neural activation in the top hidden layer $\vec{\theta}$, $f$ provides a corresponding image in reconstruction space $\vec{y}=f(\vec{\theta})$.
\label{fig:network}
}
\end{center}
\end{figure}

 The second way to view the model as a fit occurs after the network has been fully trained. Once the network is trained, we require that the weights and biases ($W$, $W'$, $b$ and $b'$ for each layer) remain fixed. The trained network provides a function $f$ which maps a vector in the top hidden layer $\vec{\theta}$ to a distinct image in the reconstruction space $\vec{y}$ such that $\vec{y}=f(\vec{\theta})$. A diagram is provided in Figure \ref{fig:network}.  The trained network provides us with a model in which a low dimensional parameter space $\vec{\theta}$ can be used to represent any image reconstructed from the dataset along with a host of others. The set of all images producible by this model forms the 'model manifold' of the trained network. As the parameters are varied, the images formed interpolate between the images the network reconstructs from the dataset. For example, examining our single-digit network, the ten neural outputs in the top hidden layer sweep out a 10 dimensional model manifold  $\{ \vec{y}\} = f(\{ \vec{\theta}\} )$ in the 784 dimensional space of possible MNIST images where the image space dimensionality is determined by the number of pixels in the images (28x28). The fit is then: given an image $\vec{d}$, vary the parameters $\vec{\theta}$ such that the cost fuction $C(\vec{\theta})=||\vec{d}-f(\vec{\theta})||^2$ is minimized.\par

 In addition to studying the model manifold corresponding to the reconstruction space, the layered structure allows for exploration of the model manifold the network has learned to represent the data in each layer of the network. Throughout,  'Layer 1'  corresponds to the reconstruction space and 'Top Layer' to the top hidden layer of the network. \par

\section{Jeffrey's Prior}
The first task in visualizing the model manifold is to generate images apart from those on which the neural network was trained. To get a picture of  what the model manifold looks like, points in the model manifold are generated using Monte Carlo sampling. To implement our sampling we use Metropolis Monte Carlo and an uninformative prior known as Jeffrey's prior. This prior is the square root of the determinant of the Fisher Information Matrix. The choice of prior was made due to the fact that it weights volume in parameter space by volume in data space. In effect, it keeps the algorithm from getting stuck in small regions of data space which correspond to a wide range of parameter values. Additionally, unlike the uniform prior, Jeffrey's prior is invariant to transformations of the parameters. \par

In order to implement Jeffrey's Prior, the metric of  the space must be defined. The mapping from the top hidden layer (parameter space) to the reconstruction space (data space) implies a natural fitting procedure for any given image. The cost for this fit is given by
\begin{equation}
	\label{eqn:cost}
	C(\theta)=\frac{1}{2}\sum_k(y^\theta_k-d_k)^2
\end{equation}
\noindent where $\vec{y}\: ^{\vec{\theta}}$ is the reconstruction with corresponding top hidden layer activations $\vec{\theta}$ and $\vec{d}$ is the data point. The Hessian is then
\begin{equation}
	\begin{split}
	H_{ij}&=\frac{\partial^2 C}{\partial\theta_i\partial\theta_j}=\frac{\partial}{\partial\theta_i}\bigg{(}\frac{\partial}{\partial\theta_j}\sum_k\frac{(y_k^\theta-d_k)^2}{2}\bigg{)}\\
	&=\sum_k\bigg{(}\frac{\partial y_k^\theta}{\partial\theta_i}\frac{\partial y_k^\theta}{\partial\theta_j}+(y_k^\theta-d_k)\frac{\partial^2y_k^\theta}{\partial\theta_i\partial\theta_j}\bigg{)}
	\end{split}
\end{equation}
The second term in this expression is computationally expensive and is exactly zero for data described by the model. Additionally, even for data points not lying on the model manifold, the values in this sum fluctuate between positive and negative, averaging to zero. These characteristics make an approximation which neglects this term very natural:
\begin{equation}
	H\approx\sum_k\bigg{(}\frac{\partial y_k^\theta}{\partial\theta_i}\frac{\partial y_k^\theta}{\partial\theta_j}\bigg{)}=J^TJ
\end{equation}
In addition to being less expensive to compute, the approximate Hessian, aka the Fisher information Matrix, is positive definite and data independent. Indeed it is the metric on the space of neural outputs (top hidden layer) induced by the least-squares metric in data (image) space. The distance between two nearby neural outputs is given by the squared difference of their corresponding images.\par

Now consider the singular value decomposition (SVD) of the Jacobian matrix, $J=U\Sigma V^T$, where $V$ is a orthogonal matrix in parameter space, $\Sigma$ is a diagonal matrix of the singular values and the columns of $U$ form an orthonormal basis in data space which span the range of J. The metric can thus be written as
\begin{equation}
	g=V\Sigma^2V^T
\end{equation}
where the columns of $V$ correspond to the eigenparameters and the eigenvalues are given by $\lambda_i=\Sigma_{ii}^2$. Geometrically, this states that the Jacobian maps metric eigenvectors into the data space vectors $U_i$ stretched by a factor $\sqrt{\lambda_i}$. The mapping from hidden layers into data space expands volume (N-volume to N-dimensional surface area) by a factor $\prod _i \sqrt{\lambda_i}= \prod_i \Sigma_i$. \par

In order to sample according to Jeffrey's prior, the points are given a prior probability that is equal to the square root of the determinant of $g_{\alpha\beta}$. So we have
\begin{equation}
p(\theta)=\sqrt{|g_{\alpha\beta}|}=\sqrt{|J^TJ|}=\prod_i \Sigma_i
\end{equation}
where $\Sigma_i$ are the singular values of the Jacobian. This probability density in the space of neural outputs thus samples surface volumes in the model manifold equally. Performing Metropolis Monte Carlo on the model manifold using Jeffrey's prior enables us to explore the model manifold of the single-digit network as well as the DBN and SdA. \par

\section{Results}
\label{sec:results}

Sampling of the model manifold of each trained neural network is performed using emcee - an MIT licensed pure-Python implementation of Goodman \& Weare’s Affine Invariant Markov chain Monte Carlo (MCMC) Ensemble sampler \cite{GoodmanWeare, emcee}. In particular, we found it useful to employ their parallel tempering package. For the single-digit network we used 6 temperatures with 20 walkers each chosen randomly from MNIST images of the digit '1'. The results presented are for a sampling of 500,000 steps for each walker at a temperature of 1.0. Similarly, for the DBN and SdA we used 6 temperatures each although the number of walkers was increased to 60 as the dimensionality of the parameter space increased three-fold. Walkers were again chosen randomly from MNIST digits. Each walker was run for 50,000 steps. Results shown are for a temperature of 1.0.\par

\begin{figure}
\begin{center}
\includegraphics[width=0.9\textwidth]{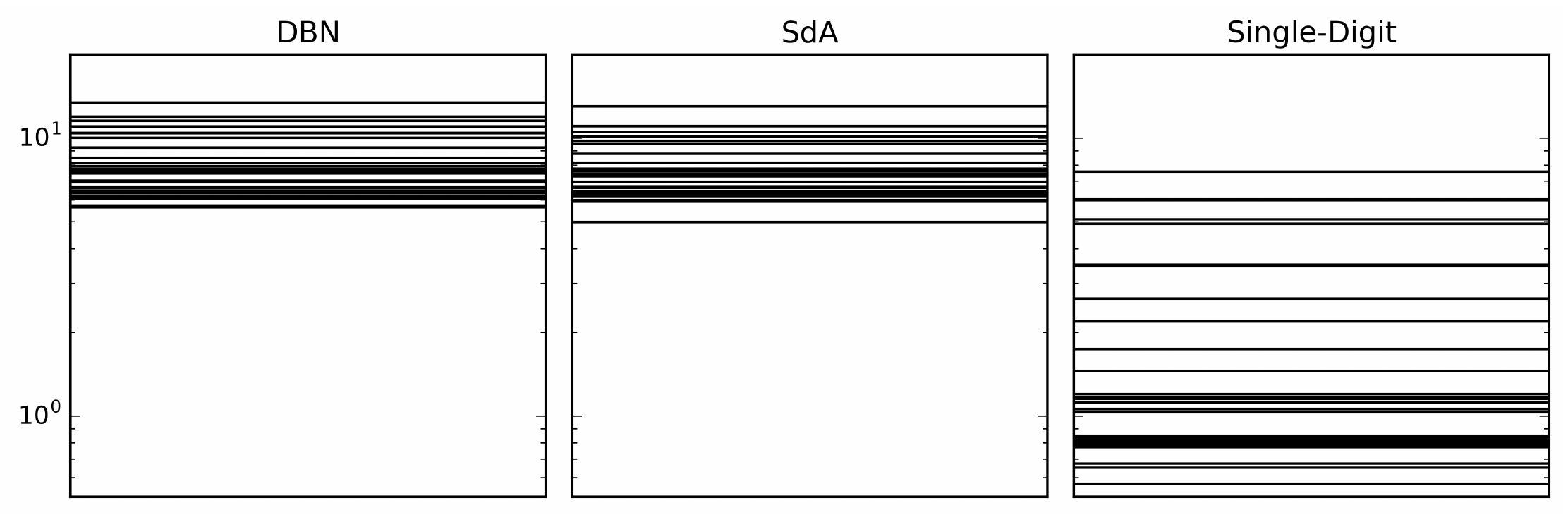}
\caption{Widths of the reconstructed manifold for each network along PCA directions.
\label{fig:widths}
}
\end{center}
\end{figure}

\begin{figure}
\begin{center}
\includegraphics[width=0.9\textwidth,trim={3cm 3.5cm 0 0},clip]{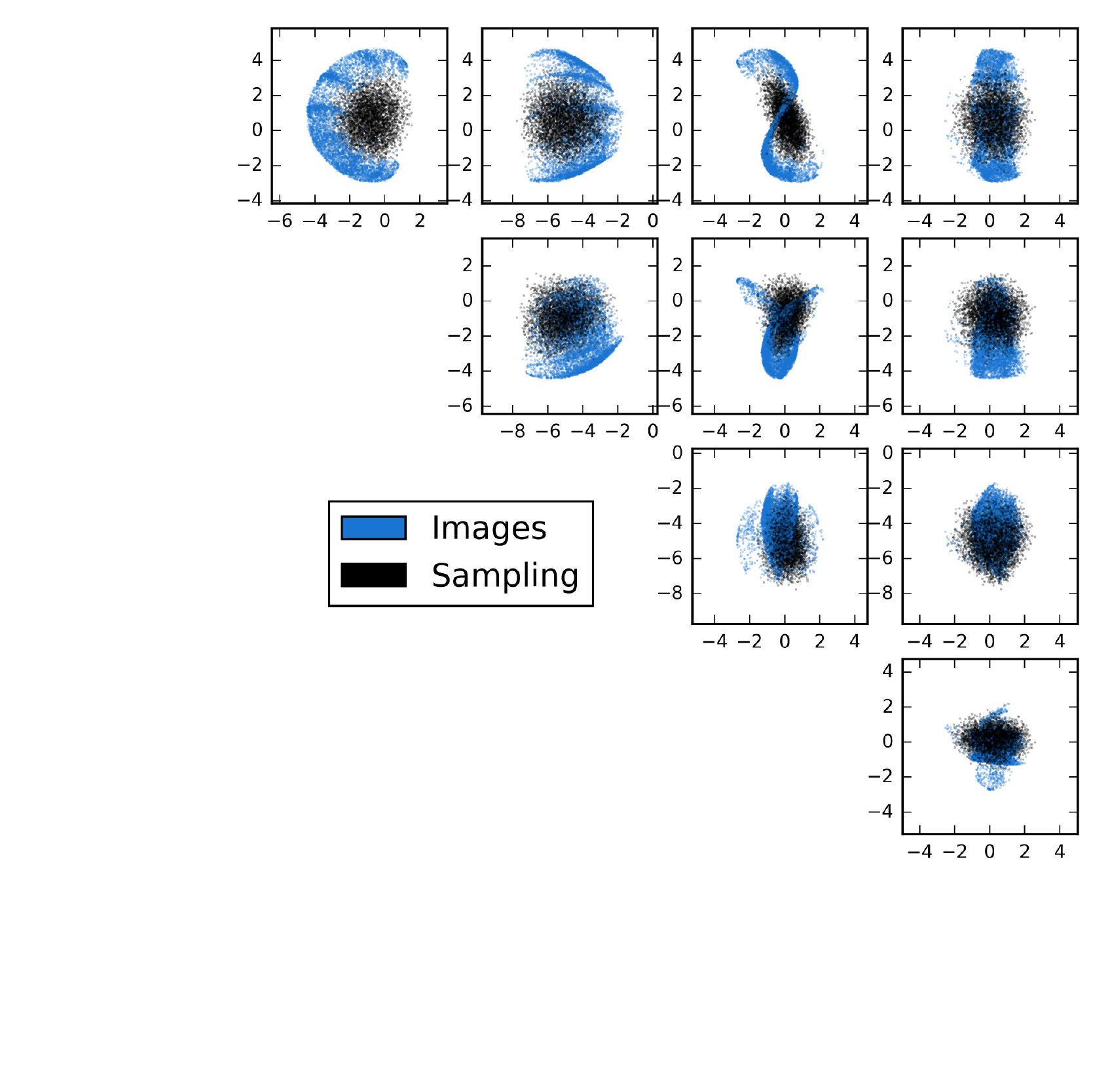}
\caption{(Color online) Two dimensional PCA projections of the MNIST digit data for ones and the Jeffrey's Prior sample of the model manifold. The $(i,j)^{th}$ figure in the grid is a plane spanned by singular vectors i and j + 1 of the centered data. Image reconstructions lie on the edges of the manifold. Width of the network decreases slightly along each principal vector; however the aspect ratio is much closer to unity than that observed in other models \cite{hyperribbon}. 
\label{fig:onesPCA}
}
\end{center}
\end{figure}

\begin{figure}
\begin{center}
\includegraphics[width=0.9\textwidth,trim={0 1cm 0 0},clip]{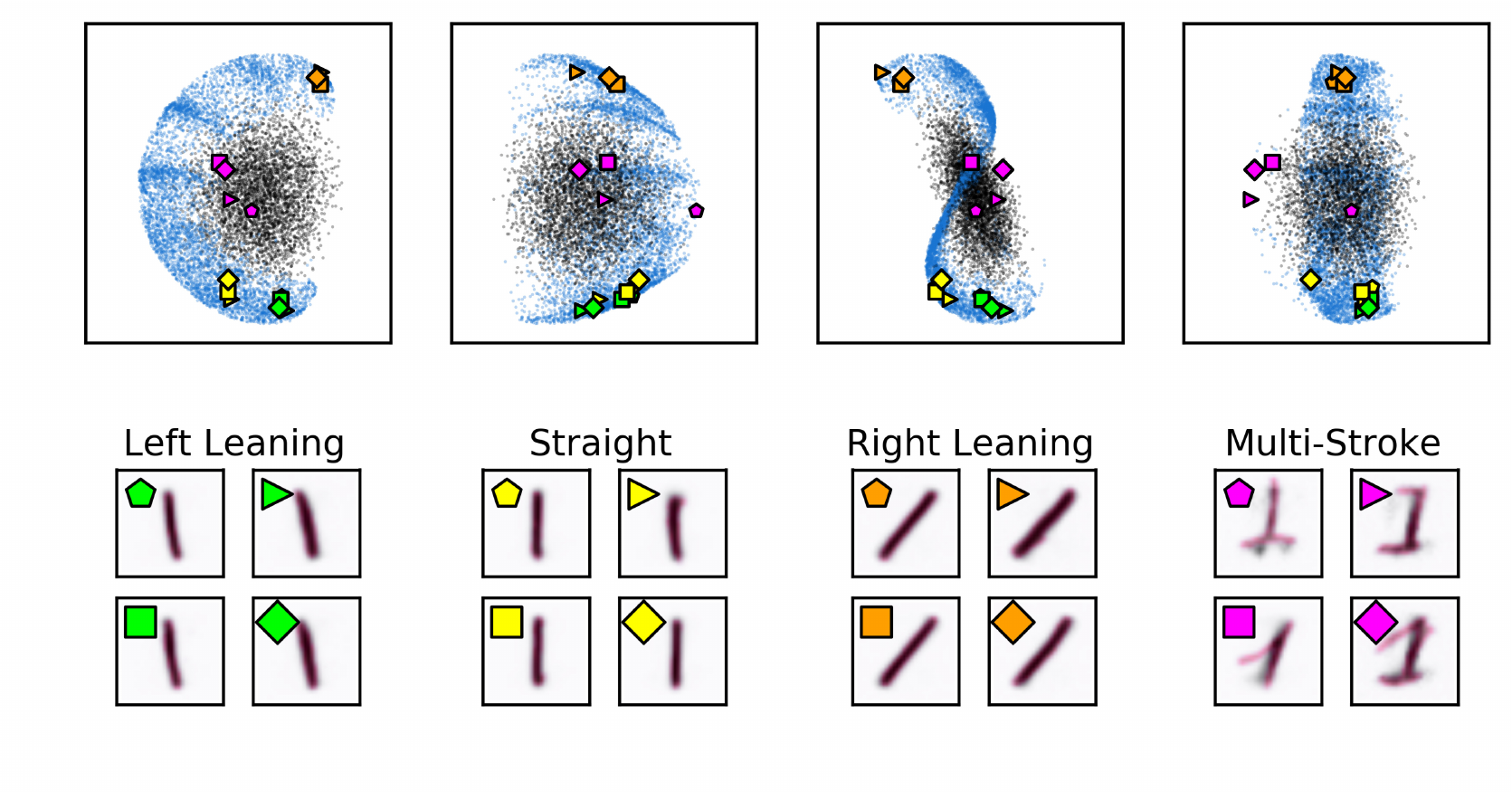}
\caption{(Color online) Two dimensional PCA projections of the MNIST digit data for ones, the Jeffrey's Prior sample of the model manifold and a few selected images from the dataset. The top row corresponds to the top row in Figure \ref{fig:onesPCA}. Ones with differing distinct characteristics such as tilt or bases have been labeled to display how the manifold is roughly arranged according to digit behavior. The images corresponding to these digits are shown on the bottom row. For each, the original digit has been plotted in pink with the reconstruction overlayed in black.
\label{fig:oneslocation}
}
\end{center}
\end{figure}

\begin{figure}
\begin{center}
\includegraphics[width=0.9\textwidth,trim={3cm 3cm 0 0},clip]{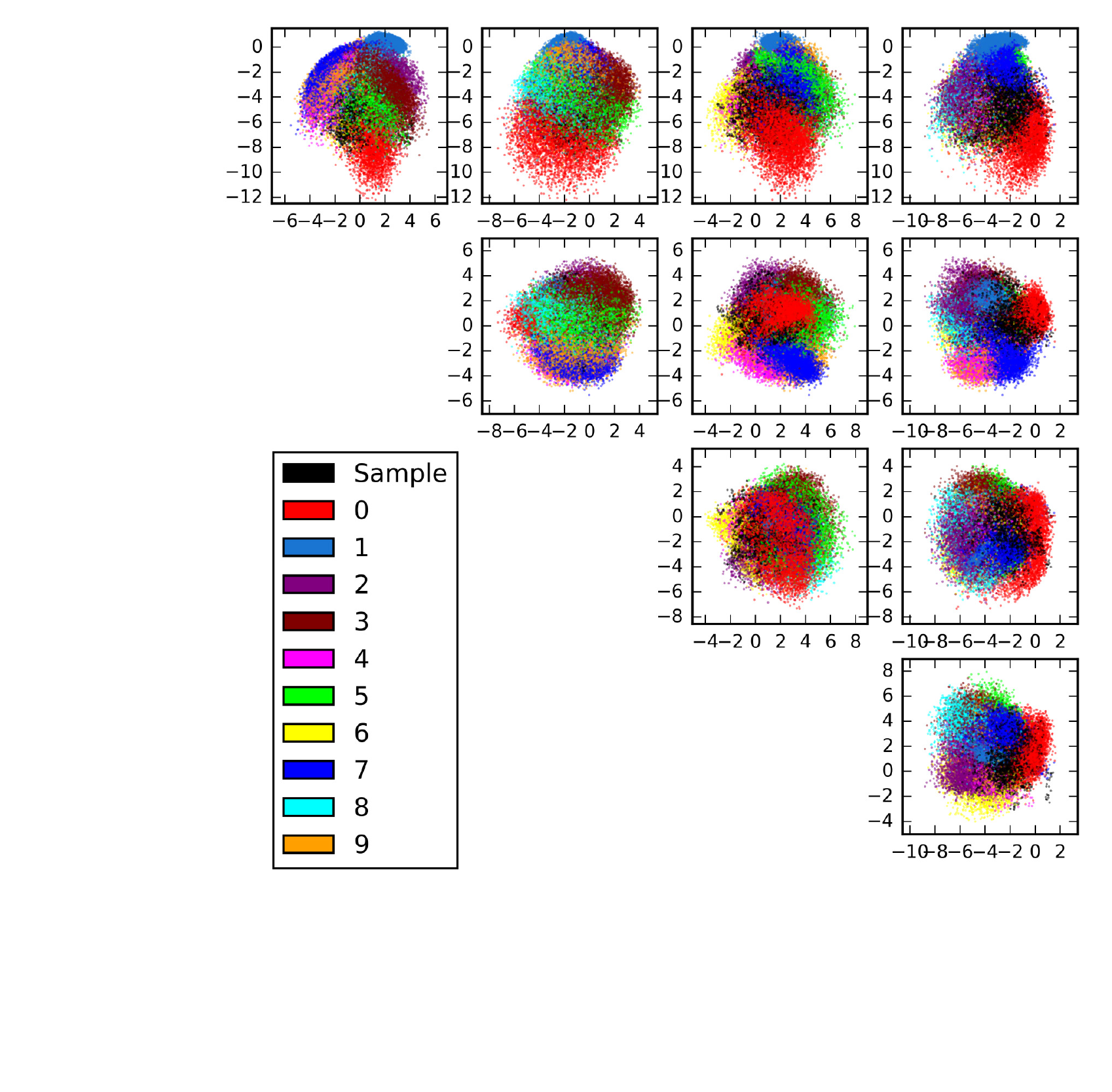}
\caption{(Color online) Two dimensional PCA projections of the MNIST digit data and the Jeffrey's Prior sample of the model manifold for a 4 layer DBN. The $(i,j)^{th}$ figure in the grid is a plane spanned by singular vectors i and j + 1 of the centered data. 
\label{fig:dbnPCA}
}
\end{center}
\end{figure}

 PCA images of the sampled points for the single-digit network are shown in Figure \ref{fig:onesPCA}. This figure corresponds to PCA projections of the data where the $(i,j)^{th}$ figure in the grid is a plane spanned by singular vectors i and j + 1 of the centered data. Presenting PCA projections in this way reveals the striking structure which led us to consider the ones. The projections along the vector corresponding to the largest principal component are shown in the first four frames along the top row.  In these images one can see that data  (reconstuctions of the MNIST digit 1's) lies in an arc. More discussion of this arc and how ones with different characteristic arrange themselves are presented in Figure \ref{fig:oneslocation}.   PCA projections of the sampling for the DBN are shown in \ref{fig:dbnPCA} with sampled digits in Figure \ref{fig:dbnwalk}. Results for the SdA displayed the same behavior and are collected in the Appendix \ref{app:neuralnet}.  In each of these images, the sampling forms a slightly elongated hyperball with corresponding hierarchy of PCA widths shown in Figure \ref{fig:widths}.

For the single-digit network, the top hidden layer has dimension 10 hence the reconstructions must be a 10 dimensional object embedded in the 784 dimensional pixel space. We find that in each successive PCA direction, the width of the manifold shrinks by a factor of 1.17 along the first 10 PCA vectors.  For the DBN and SdA, the reconstructed manifold is 30 dimensional and the factor is a meager 1.03. In addition to PCA, there were other methods of determining widths we explored detailed in Section \ref{sec:discussion}. These factors clearly contrast the structure of the neural network model manifold with the hyper-ribbon manifolds observed in sloppy models. 

\begin{figure}
\begin{center}
\includegraphics[width=0.9\textwidth]{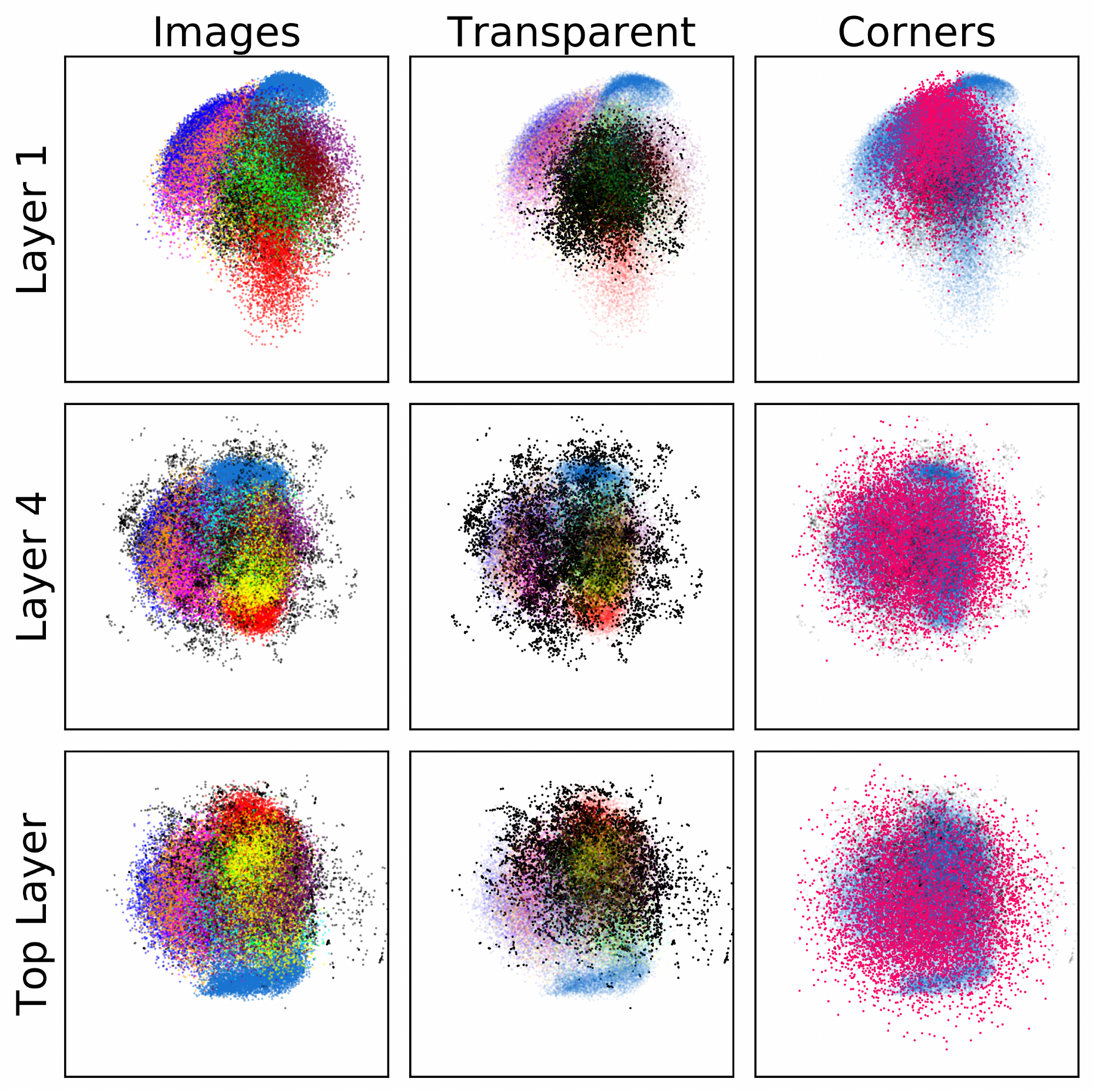}
\caption{(Color online) PCA projection of the Jeffrey's Prior sampling with MNIST digit data for the DBN along the two largest principal component vectors. In the middle column 'Transparent' the transparency of the MNIST points have been enhanced to show the position of the sampling. On the right under 'Corners' the digits and sampling have again been plotted in blue and black respectively with the corners added in pink. Corners correspond to activations in the top hidden layer ($\vec{\theta}$) for which $\theta_i \in\lbrace -\infty,\infty \rbrace \sim \lbrace -10^6, 10^6 \rbrace$. The axes are shared along each row. In order to deal with the large values of the corners and sampling, the top layer is shown with a sigmoid applied. Although the sampling spans the parameter space (Top Layer), it is subsequently mapped to the interior of the manifold in reconstruction space ('Layer 1').  Note that the radius of the sampling is roughly $\frac{2}{3}$ that of the digits - resulting in a 30-D volume $\big{(}\frac{2}{3}\big{)}^{30}=5.2\times10^{-6}$ of the total volume. Notice also that the digits lie outside the corners.
\label{fig:dbnall}
}
\end{center}
\end{figure}

Another characteristic of each PCA projection plot is that, in each, points representing MNIST data are located along the edges of the model manifold while the bulk of the sampling remains deep in the interior. This is made more evident in  'Transparent' column of Figures \ref{fig:dbnall} and \ref{fig:sdaall} by increasing the transparency of the points corresponding to MNIST digits. Examining higher layers of each network reveals that the behavior of the single-digit network remains unchanged. For the DBN and SdA, however, as we progress to higher layers, the behavior becomes reversed. Instead of the sampling lying deep in the interior, the sampling spans a much larger distance in parameter space. We believe this to be due to a linear top layer in the SdA and DBN in contrast to the single-digit network in which all layers were sigmoidal. \par

In addition to sampling, we also explore the location of the 'corners' in relation to the images and sampling. Corners correspond to datapoints for which the top hidden layer representation is a vector with each element at an extremum allowed by the network. More specifically, in the single-digit network, each parameter in the top hidden layer is contrained to lie between 0 and 1. Using this we can plot the representation for these $2^{10}$ corner points.  For the DBN and SdA, the top layer activation is linear hence we set 'corners' to be those hidden layer activations for which $\vec{\theta}\in \lbrace-10^6,10^6\rbrace$. The top layer of each of these networks has a dimension of 30, hence there are $2^{30}$  such corners. For this reason, only a random subset of 10,000 are plotted. The PCA projections for these are in the 'Corners' columns of Figures \ref{fig:dbnall}, \ref{fig:sdaall} and \ref{fig:onesall}. By definition the corners are extremely far apart in the top layer. For this reason each plot of the top layer is presented with a sigmoid applied. For each of the 4 layer networks the nonlinearity of the network results in the interior of the space bulging out such that the corners lie within the volume of the manifold in data space (Layer 1) rather than on the boundaries. \par

The 'digits' sampled by each of these networks do not correspond to actual images. Several of these sampled images for the DBN are shown along with the original images and 'eigen-images' of the dataset in Figure \ref{fig:dbnwalk}. The eigen-images are formed by taking the singular value decomposition of the MNIST dataset such that 

\begin{equation}
X = U\Sigma V^T 
\end{equation}

\begin{figure}
\begin{center}
\includegraphics[width=0.9\textwidth]{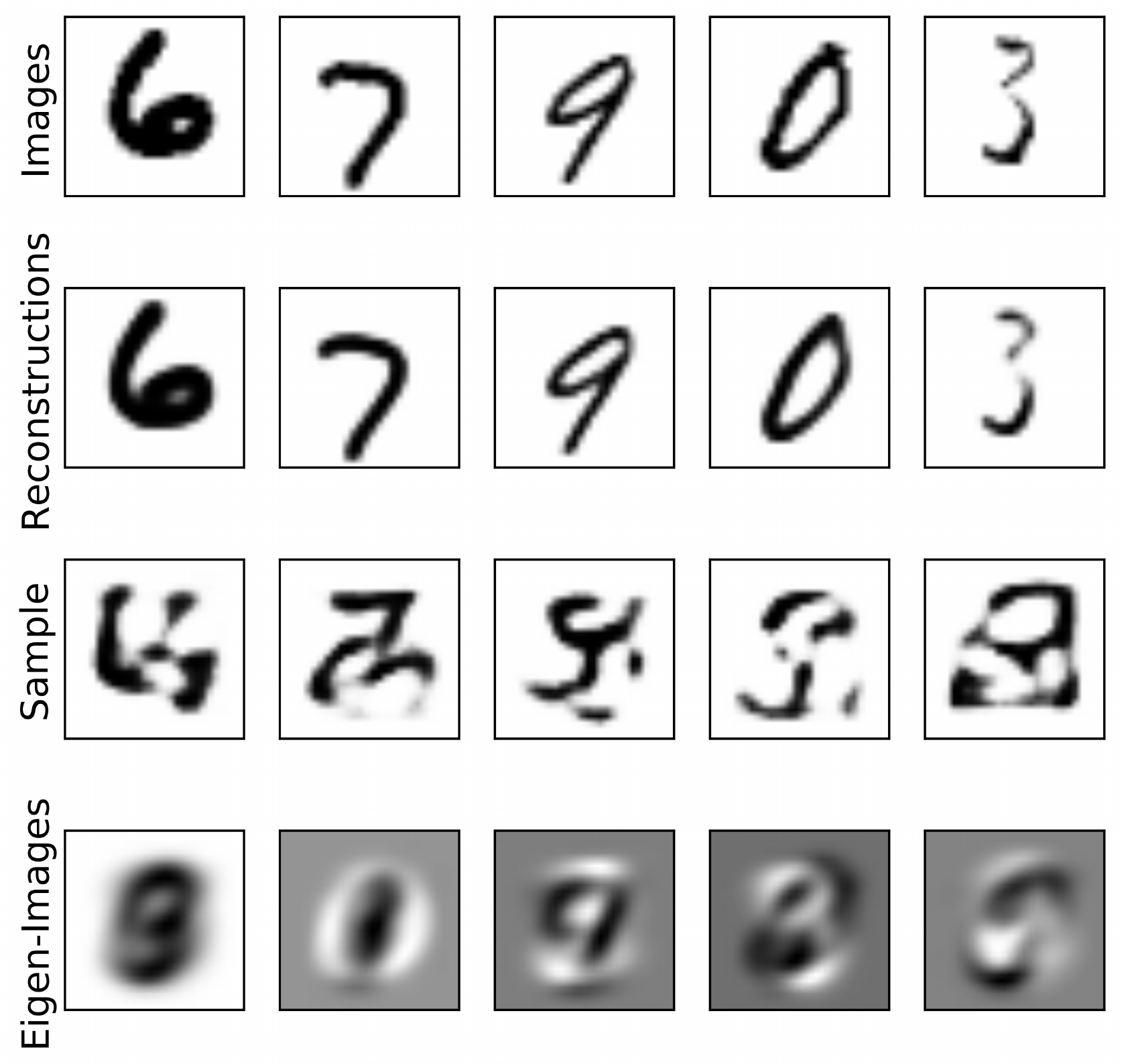}
\caption{Examples of MNIST images, their reconstructions, and images sampled using Jeffrey's Prior for the DBN. For the sampled 'digits', each snapshot corresponds to the same walker. The final row corresponds to the top five 'eigen-digits' of the dataset.
\label{fig:dbnwalk}
}
\end{center}
\end{figure}

\noindent where $X$ is the data arranged as (number of samples) x (image size). The images in the  matrix $V$ which correspond to the eigenbasis of $X$ compose the eigen-images. For the single-digit network and SdA network, the corresponding plots are found in the Appendix \ref{app:neuralnet}. For each network, although the eigen-images do not form actual images, they show more identifiable structure than the sampled images. Additionally, they are located on the boundary of the manifold in reconstruction space just as are the actual images. This characteristic location provides further indication that the boundaries play an important role in the model.\par

\section{Discussion}
 \label{sec:discussion}

A deep neural network with N hidden outputs provides a N dimensional representation of a high dimensional data set. As we vary N, how does the resulting hierarchy of descriptions reflect itself in the geometry of these models? Our hypothesis was that the neural manifolds would form a hyper-ribbon, with incremental expressiveness as $N \rightarrow N-1$ reflected in geometrically thinner geodesic widths. This hypothesis, verified for multiparameter models in other disciplines \cite{hyperribbon} appears not to be the case for the neural manifolds we study. Instead, their model manifold forms an only slightly elongated hyperball. This is striking in that, for a typical model, such behavior would imply that the model tends to weight parameter combinations more or less equally. In other words, for the neural network, it would imply that there are no neurons which, in tandem, control the majority of behavior. Conversely, there are no sets whose values can change dramatically without affecting the output. \par  

There are several techniques that have been found to be useful for training of neural networks; purportedly due to keeping the network from relying too heavily on any one neuron or set of neurons. For example, the L1 and L2 norm terms penalize large weights which would allow certain neurons to swamp the signal. Other techniques such as dropout \cite{dropout} explicity drop neurons at random during training forcing all neurons to 'pull their own weight'. Another successful technique, channel out \cite{channel1,channel2}, allows sets of neurons to be activated only for certain tasks. The goal, however, is still to make use of the whole network and avoid computational waste. Yet other techniques employ initializations designed to prevent the network from collapsing onto a few modes \cite{orthogonal}. In the case presented here, only simple whitening of the data was used. \par

In addition to the PCA plots, we employed a number of other tactics to search for a thin direction. These include sampling of slices of the manifold and a geodesic analysis. For the former, we found the two furthest points in the manifold and defined a  slice at the midpoint of this vector. Sampling was then performed in the slice and the procedure repeated. This procedure yields a sequence of slices which we constrained to be perpendicular. The hierarchy of widths uncovered by this procedure did not yield a thin direction. In addition, several variants were implemented in which we sliced along the densest section rather than midpoint and varied the width of the slice. We also explored the widths using geodesic analysis. Calculating geodesics from a central point radiating in a plane can be used to map the boundaries. We found a suggestion of thin directions. Under further scrutiny, these appeared to be due to curvature which caused the geodesic to strike a boundary prematurely. In all of our investigations, the manifold truely seems to form an only slightly elongated hyperball rather than a hyper-ribbon. \par

There is a characteristic of the dataset and networks we employ, however, which obscure the clear geometrical arguments made in other fields for the relative importance of parameter combinations. Namely, the data itself forms a hyperball. All images of digits have many saturated (white/black) pixels corresponding to the boundary of the manifold as dictated by the sigmoidal structure of the SdA and DBN. For this reason, it is difficult to state with certainty that the structure observed can be interpreted in the standard way.\par

Instead,  we find evidence motivating a hypothesis that the N-1 and N-dimensional descriptions have a boundary relationship: $\mathcal{M}_{N-1} \approx \partial\mathcal{M}_N.$ First, we find that the reconstructed data lies on the 'outside' of the neural manifold (Figures \ref{fig:dbnall}, \ref{fig:sdaall} and \ref{fig:onesall}), forming a shell around the $\mathbb{L}^2$ Jeffrey's prior volume of the possible image reconstructions. This naturally suggests that lower dimensional representations will also hug the outside. Second, we have explicitly constructed lower dimensional neural manifolds by training an additional autoencoder layer on the 30 dimensional top hidden layer of the DBN and SdA. In each case, we first applied a sigmoid and the hidden layer consisted of 3 neurons. PCA projections of the corresponding 3 dimensional model manifold along with the original 30 dimensional version embedded in data space are shown in Figures \ref{fig:DBN3D} and \ref{fig:SdA3D}.  For the DBN  the lower dimensional representation is consistently on the boundary indicating that $\mathcal{M}_3 \stackrel{\subset}\sim ... \stackrel{\subset}\sim\partial \mathcal{M}_{29} \stackrel{\subset}\sim \partial \mathcal{M}_{30}$. For the SdA, PCA along the first two principal components displays the lower dimensional manifold on the boundary. In the third, it appears to pierce the space. The emphasis on the boundaries for deep networks is in parallel to behavior observed in other models. Recently, work by Transtrum et. al. \cite{Transtrum} has explored the relationship between manifold boundaries and emergent model classes. Their work in information topology has uncovered that boundaries of the model manifolds they explore each correspond to different classes of reduced models. The topology of the space in turn governs the best low dimensional description. \par

%
%
%
%
%
%

Geometry is strange in high dimensions. This can be illustrated by the following apparent contradiction: the majority of the volume of a hyperball is located near the boundary but it takes a huge sampling to obtain even one point in a cap of the ball. Given our observation that $\mathcal{M}_{30}$ is roughly a slightly elongated hyperball, 95.8 \% of the the volume is within 10\% of the surface. This appears to imply that Jeffrey's prior should be appropriate to sample the region near the boundary on which the actual reconstructed data lies. Conversely, consider a N dimensional hyperball of radius, r, with a cap of height, h. The volume of the cap is given by

\begin{equation}
V_n^{cap}(r) = \frac{1}{2}V_n(r)I_{sin^2\phi}\bigg{(}\frac{n+1}{2},\frac{1}{2}\bigg{)}
\end{equation}

\noindent where $V_N$ is the volume of the hyperball, $I$ corresponds to the regularized incomplete beta function and $\phi = sin^{-1}\frac{r-h}{r}$ is the colatitude angle such that $0\leq\phi\leq \frac{\pi}{2}$ \cite{hypersphere}. For a 10-D sphere with radius 1 and cap height 0.1, the volume of the cap corresponds to a tiny  0.0014\% of the volume of the ball. In terms of sampling, this states that one would need to sample approximately $10^5$ points with Jeffrey's Prior to obtain even a single point in this cap. In this light, it seems that instead, data located on the boundary in the way we observe suggests that a prior based on volume actually biases against the interesting regions.

The model manifolds formed by the neural networks we study have novel properties. In short, they do not appear to form the hyper-ribbons seen in other fields. Additionally, for Deep Belief Networks and Stacked Denoising Autoencoders trained on the MNIST digit dataset, the vast majority of the data arranges itself of the boundaries of the manifold. Sampling uncovers that the interior, which represents the bulk of the images the network can describe, does not correspond to actual images. This implies that either the network is in some sense wasteful or that Jeffrey's Prior is naturally biased to explore uninteresting regions of the model. \par


%

\appendix*

\section{Additional Results}\label{app:neuralnet}

\begin{figure*}
\begin{center}
\includegraphics[width=0.9\textwidth]{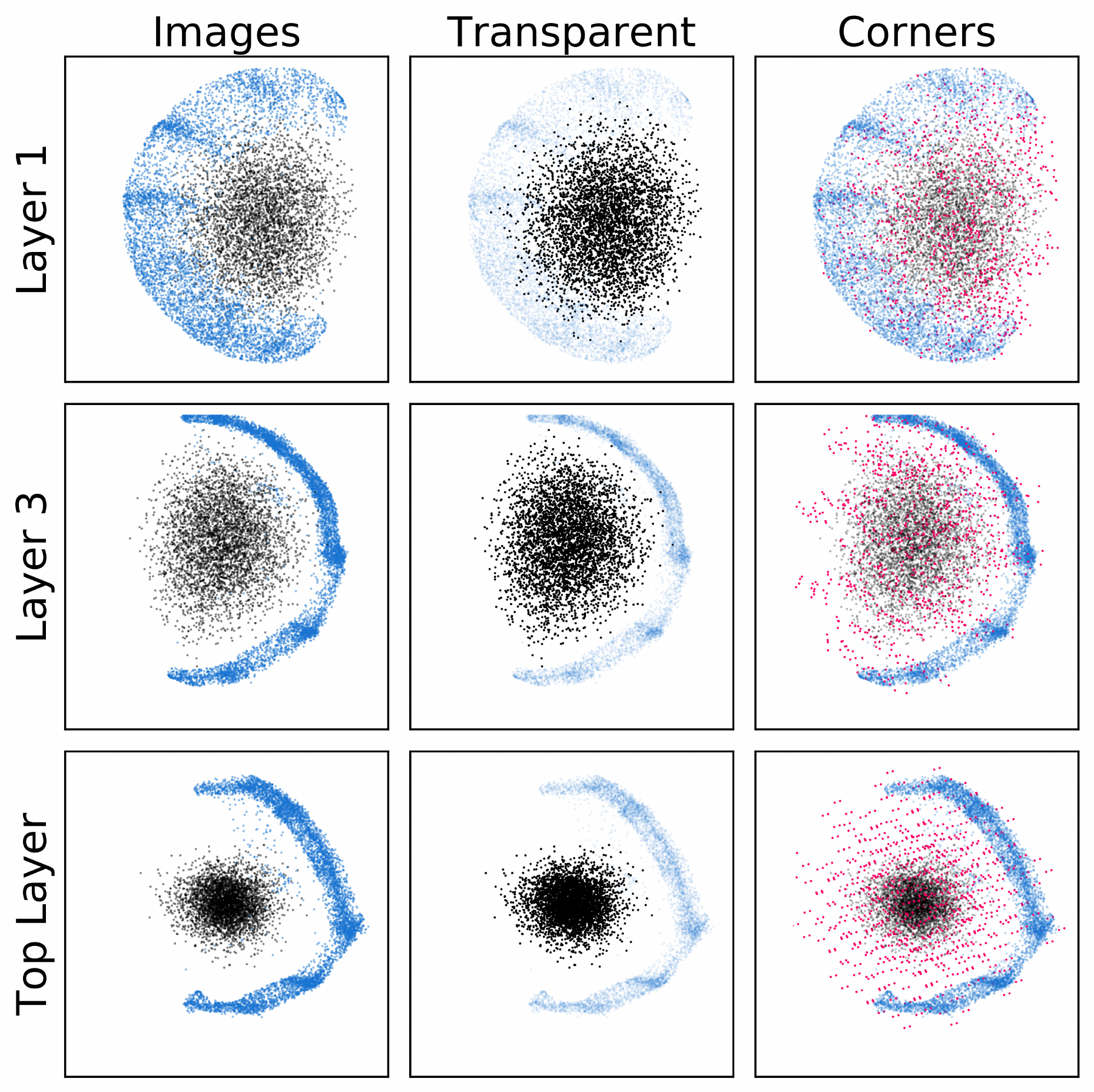}
\caption{(Color online) PCA projection of the Jeffrey's Prior sampling with MNIST '1' digit data for the single-digit network along the two largest principal component vectors. In the middle column 'Transparent'  the transparency of the MNIST points have been enhanced to show the position of the sampling. On the right under 'Corners' the '1's and sampling have again  been plotted in blue and black respectively with the corners added in pink. Corners correspond to activations in the top hidden layer ($\vec{\theta}$) for which $\theta_i \in\lbrace 0, 1 \rbrace$. 
\label{fig:onesall}
}
\end{center}
\end{figure*}

\begin{figure*}
\begin{center}
\includegraphics[width=0.9\textwidth]{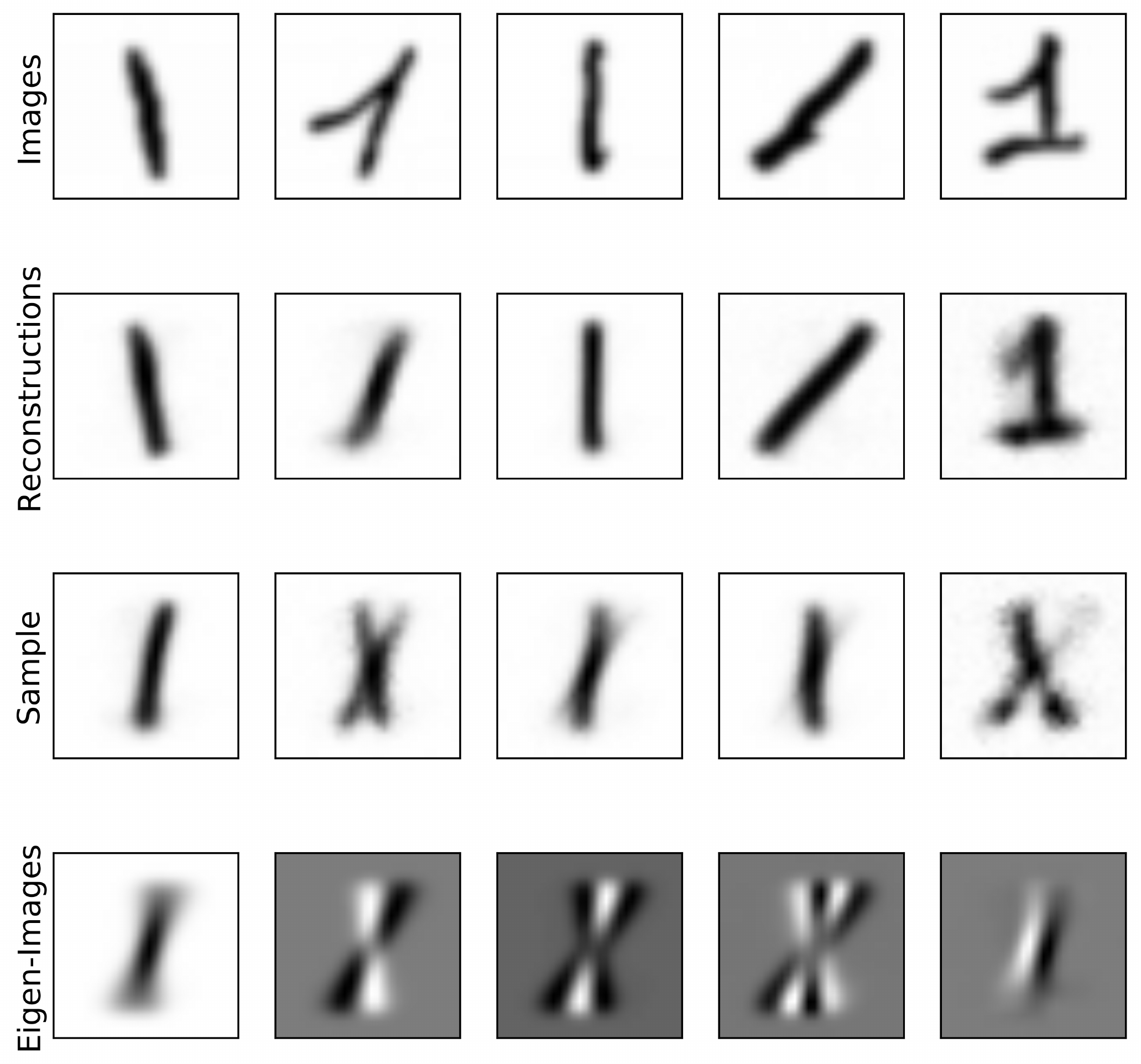}
\caption{(Color online) Examples of MNIST '1' images, their reconstructions, and images sampled using Jeffrey's Prior for the single-digit network. For the sampled '1's, each snapshot corresponds to the same walker. The final row corresponds to the top five 'eigen-digits' of the '1's dataset.
\label{fig:oneswalk}
}
\end{center}
\end{figure*}

\begin{figure*}
\begin{center}
\includegraphics[width=0.9\textwidth]{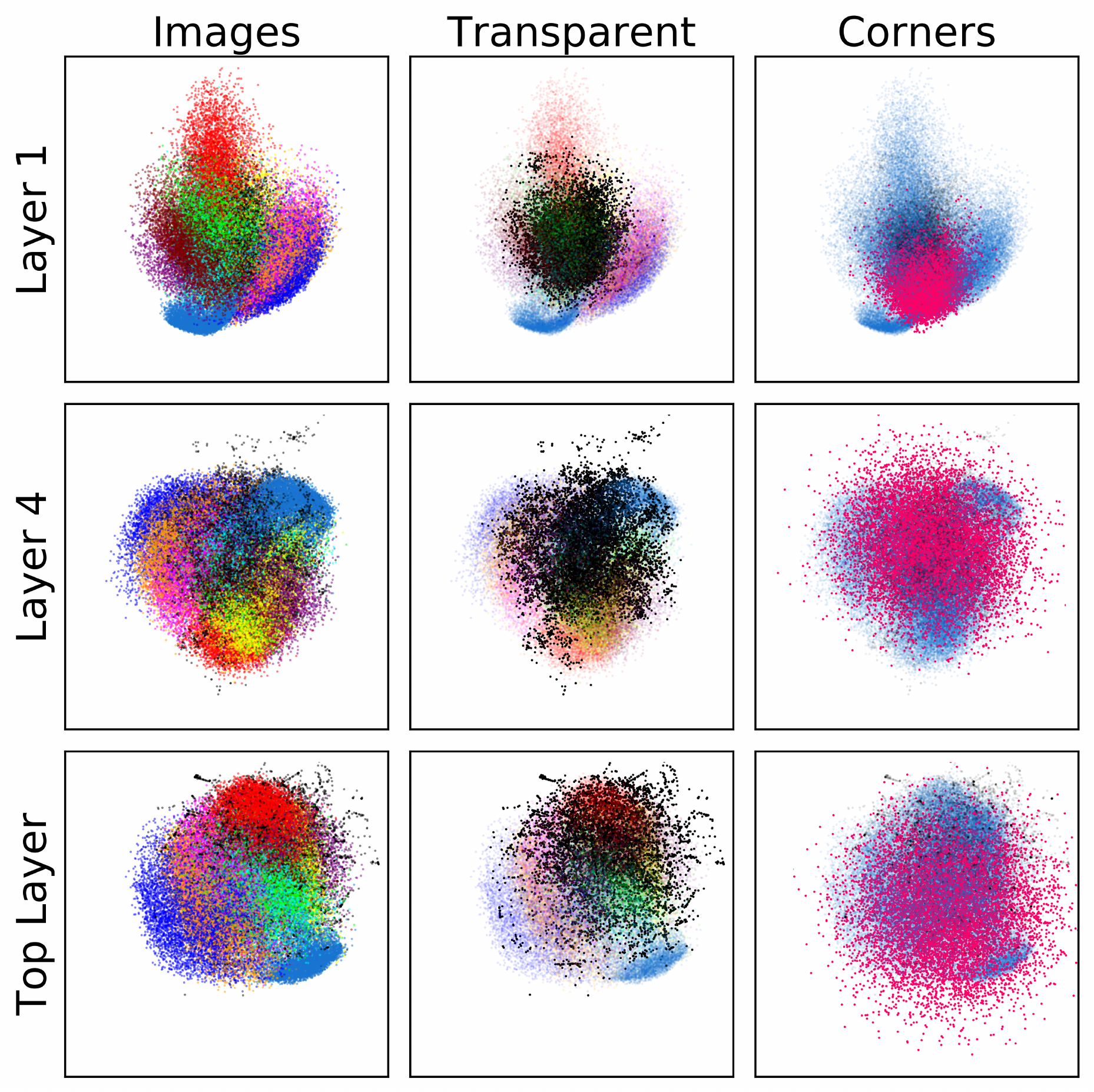}
\caption{(Color online) PCA projection of the Jeffrey's Prior sampling with MNIST digit data for the SdA along the two largest principal component vectors. In the middle column 'Transparent' the transparency of the MNIST points have been enhanced to show the position of the sampling. On the right under 'Corners' the digits and sampling have again been plotted in blue and black respectively with the corners added in pink. Corners correspond to activations in the top hidden layer ($\vec{\theta}$) for which $\theta_i \in\lbrace -\infty,\infty \rbrace \sim \lbrace -10^6, 10^6 \rbrace$. The axes are shared along each row. In order to deal with the large values of the corners and sampling, the top layer is shown with a sigmoid applied. Although the sampling spans the parameter space (Top Layer), it is subsequently mapped to the interior of the manifold in reconstruction space ('Layer 1').  
\label{fig:sdaall}
}
\end{center}
\end{figure*}

\begin{figure*}
\begin{center}
\includegraphics[width=0.9\textwidth]{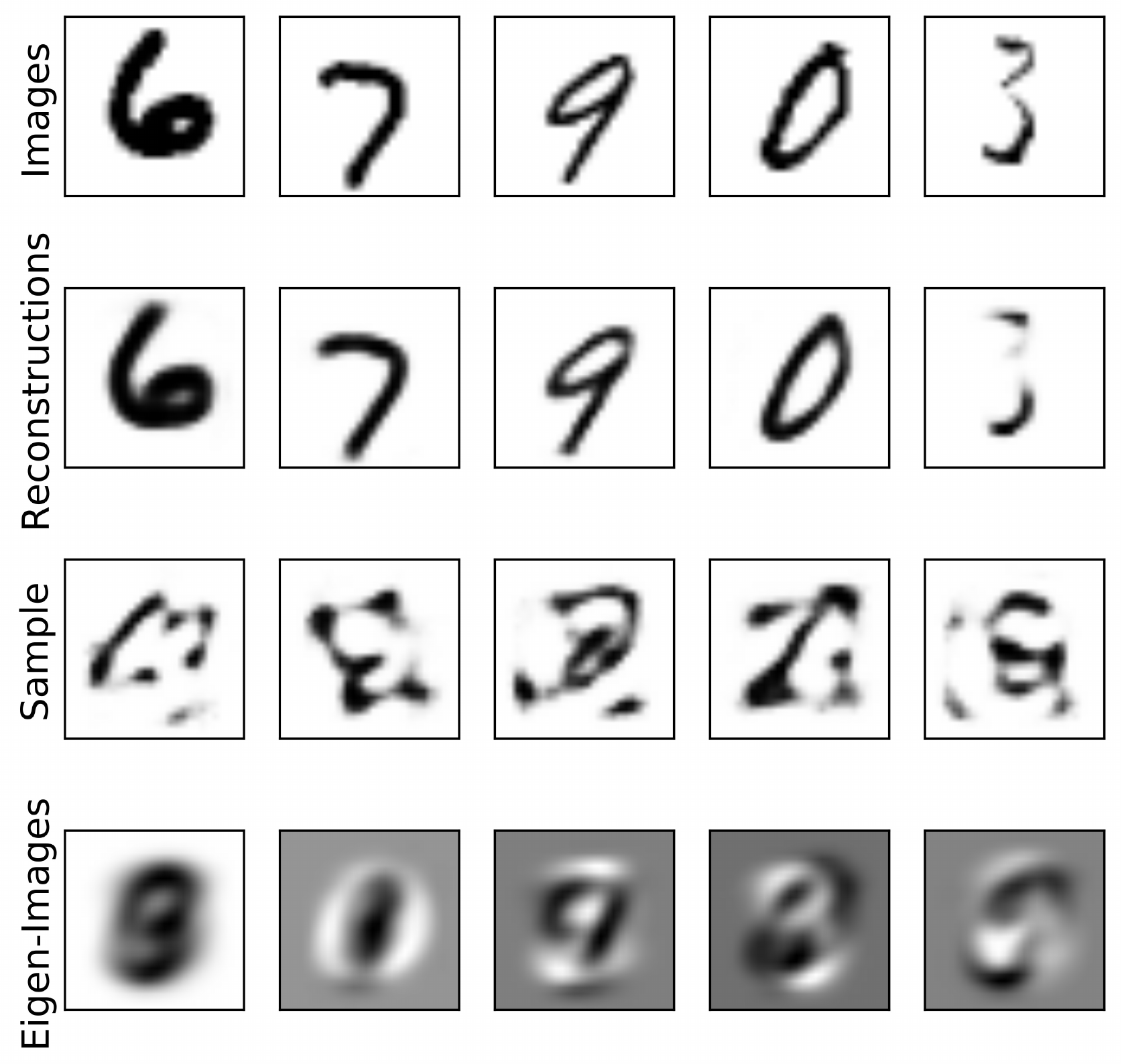}
\caption{(Color online) Examples of MNIST images, their reconstructions, and images sampled using Jeffrey's Prior for the SdA. For the sampled 'digits', each snapshot corresponds to the same walker. The final row corresponds to the top five 'eigen-digits' of the dataset.
\label{fig:sdawalk}
}
\end{center}
\end{figure*}

\begin{figure*}
\begin{center}
\includegraphics[width=0.9\textwidth,trim={4.5cm 4.25cm 0 0},clip]{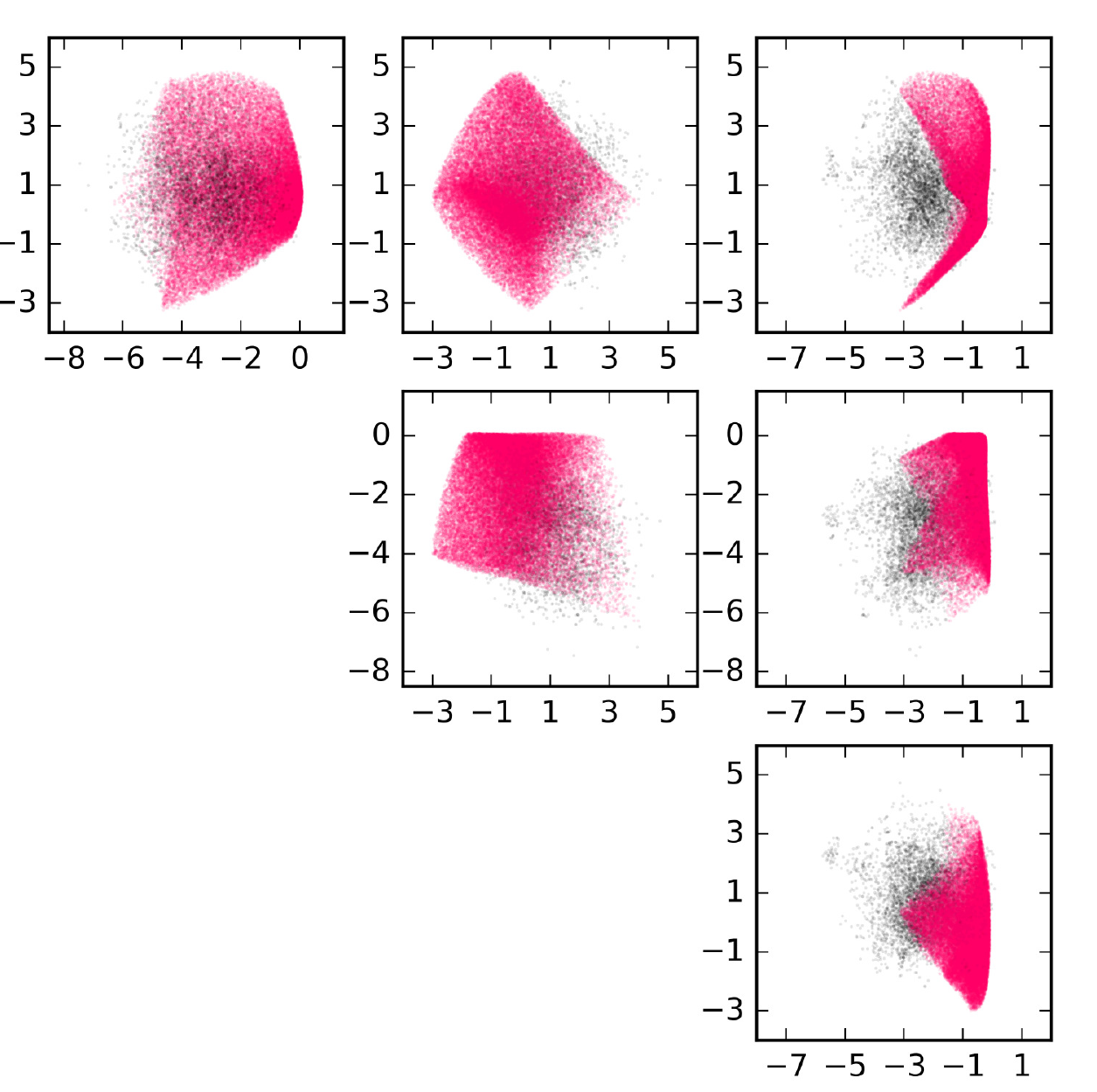}
\caption{(Color online) PCA projection of the Jeffrey's Prior sampling for the 3D manifold (Pink) and 30D manifold (Black) in reconstruction space for the DBN.
\label{fig:DBN3D}
}
\end{center}
\end{figure*}

\begin{figure*}
\begin{center}
\includegraphics[width=0.9\textwidth,trim={4.5cm 4.25cm 0 0},clip]{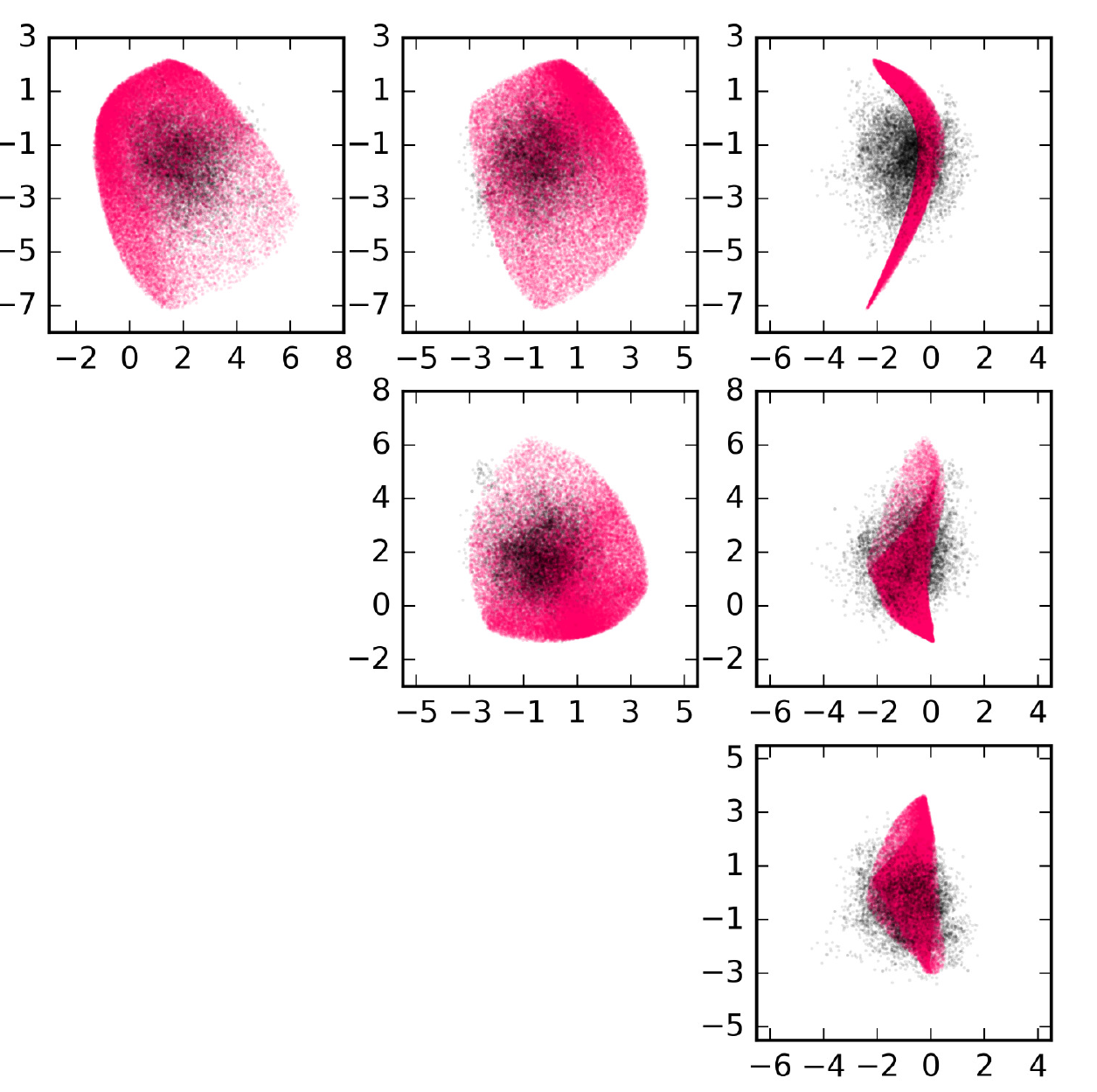}
\caption{(Color online) PCA projection of the Jeffrey's Prior sampling for the 3D manifold (Pink) and 30D manifold (Black) in reconstruction space for the SdA.
\label{fig:SdA3D}
}
\end{center}
\end{figure*}

This appendix contains the results for the single-digit and SdA networks which were left out of the main text. 

\end{document}